\documentclass[10pt,conference]{IEEEtran}

\usepackage{cite}

\ifCLASSINFOpdf
   \usepackage[pdftex]{graphicx}
\else
   \usepackage[dvips]{graphicx}
\fi
\usepackage{subcaption}

\usepackage{xcolor}

\usepackage{booktabs}
\usepackage{enumitem}
\usepackage{soul} 
\usepackage{placeins} 

\usepackage{xurl}

\usepackage[cmex10]{amsmath}
\usepackage{siunitx}
\DeclareSIUnit \voltampere { VA } 
\DeclareSIUnit \pu { pu } 
\usepackage{mathtools}
\usepackage{nicefrac}

\usepackage{array}

\ifCLASSOPTIONcompsoc
 \usepackage[caption=false,font=normalsize,labelfont=sf,textfont=sf]{subfig}
\else
 \usepackage[caption=false,font=footnotesize]{subfig}
\fi

\usepackage{float}

\usepackage{fancyhdr}

\fancypagestyle{plain}{%
    \fancyhf{}
    \fancyfoot[L]{%
        \footnotesize
        \copyright 2026 IEEE. Personal use of this material is permitted. Permission from IEEE must be obtained for all other uses, in any current or future media, including reprinting/republishing this material for advertising or promotional purposes, creating new collective works, for resale or redistribution to servers or lists, or reuse of any copyrighted component of this work in other works.
    }

}

\newcommand{\subm}[2]{\text{#1}_{\text{#2}}}
\newcommand{\myCircuit}{V2T circuit }
\newcommand{\myCircuitns}{V2T circuit}

\newcommand\mydots{\hbox to 1em{.\hss.\hss.}}
\newcommand{\PLH}{{\mkern-2mu\times\mkern-2mu}}

\begin{document}
%
\title{A Time-Based Readout for Vector-Matrix Multiplication in Fully Analog Memristive SNNs}





\author{\IEEEauthorblockN{Elia Mateu-Barriendos, Álvaro Gómez-Pau, Josep Rius, Daniel Arumí, Rosa Rodríguez-Montañés, Salvador Manich}
\IEEEauthorblockA{Universitat Politècnica de Catalunya - BarcelonaTech (UPC)}}


%


\maketitle
\thispagestyle{plain}

\begin{abstract}
Artificial neural networks rely on vector–matrix multiplications (VMMs), whose implementation in von Neumann architectures is dominated by costly data movement between memory and processing units. Spiking neural networks (SNNs) mitigate this bottleneck by performing in-memory, analog VMMs using memristive crossbar arrays. However, conventional current-mode readout circuits incur significant area and power overhead.

This work proposes a fully analog readout architecture based on voltage-to-time conversion of the VMM output. By sensing the column voltage, the proposed approach avoids current-mode summing and scaling circuitry, improving area and energy efficiency. Post-layout simulations of a $10\PLH1$  SNN implemented in a \qty{130}{\nano\meter} CMOS technology validate the proposed architecture, while application to a trained $64\PLH10$ SNN for digit classification further demonstrates its feasibility for SNN inference.

\end{abstract}


%
\IEEEpeerreviewmaketitle


\section{Introduction}

Inspired by the human brain, neuromorphic computing aims to emulate the event-driven and distributed processing of biological neural networks. In this paradigm, spiking neural networks (SNNs) encode information as discrete events in time (spikes), enabling asynchronous operation and improved energy efficiency when implemented in hardware \cite{Indiveri2011,Sebastian2020}. 

\begin{figure}[ht]
\centering
\includegraphics[width=0.96\columnwidth]{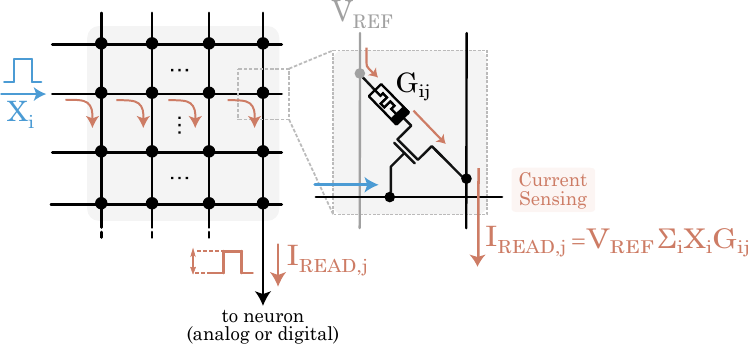}
\captionsetup{belowskip=-5pt}
\caption{An SNN with 1T1R memristive synapses employing current-mode crossbar sensing. The VMM output is represented by a column current $\subm{I}{READ,j}$.} 
\label{fig:vmm-current} 
\end{figure}

Crossbar arrays are a promising architecture for hardware SNNs \cite{Zhang2020}, where synaptic weights are stored as conductances of non-volatile memory (NVM) devices. By exploiting Ohm’s and Kirchhoff’s laws, they inherently perform vector–matrix multiplications (VMMs), typically sensed as column currents (Fig.~\ref{fig:vmm-current}). Memristive devices are attractive NVM candidates due to non-volatility, multilevel programmability, and CMOS compatibility, and are commonly used in 1T1R configurations.

While crossbar arrays enable compact in-memory VMMs, overall area and energy efficiency depends strongly on the neuron and readout design \cite{Joubert2012}. Digital neuron approaches require per-column transimpedance amplifiers and ADCs \cite{Cai2019}, leading to high area and energy consumption, or time-multiplexed ADCs \cite{Xiao2022}, which reduce
area at the cost of throughput. 

Analog neurons integrate column current on a capacitor, enabling low-power operation. However, scalability is limited by the large read currents. Metal-oxide memristive devices typically operate in the tens to hundreds of \unit{\kilo\ohm} \cite{Sun2018}, resulting in column currents in the \unit{\micro\ampere}$-$\unit{\milli\ampere} range for large arrays. \textcolor{black}{Obtaining appropriate integration times} then requires either large capacitances or additional readout circuitry, or restricting synaptic weights to binary values \cite{Valentian2022}.

Conventional solutions rely on opamp-based readout \cite{Wu2015} and current scaling circuits \cite{Garg2024}, which may also be required in digital schemes \cite{Cai2019, Jiang2023}. However, these circuits often dominate area and power, and large scaling factors over wide dynamic ranges require complex multi-stage designs \cite{Fierro2023}.

These limitations motivate alternatives to current-mode sensing. Voltage-sensing schemes have been used in compute-in-memory systems \cite{Wan2020}, while time-based encoding has been implemented in digital architectures \cite{Marinella2018,Boro2025}. However, these approaches are not directly suited to analog SNNs.

This work proposes a readout circuit based on a voltage-to-time conversion of the VMM output for fully analog SNN. By avoiding current-mode readout and explicit scaling circuitry, the approach enables compact neuron design, multi-level weights, and improved area and energy efficiency.

This paper is organized as follows. Section~\ref{sec:delay-circuit} presents the proposed architecture. Section~\ref{sec:results} describes the \(10 \PLH 1\) SNN prototype in \qty{130}{\nano\metre} CMOS and reports post-layout results.


\section{Voltage-to-time readout architecture}
\label{sec:delay-circuit}

\begin{figure}[ht]
\centering
\includegraphics[width=0.93\columnwidth]{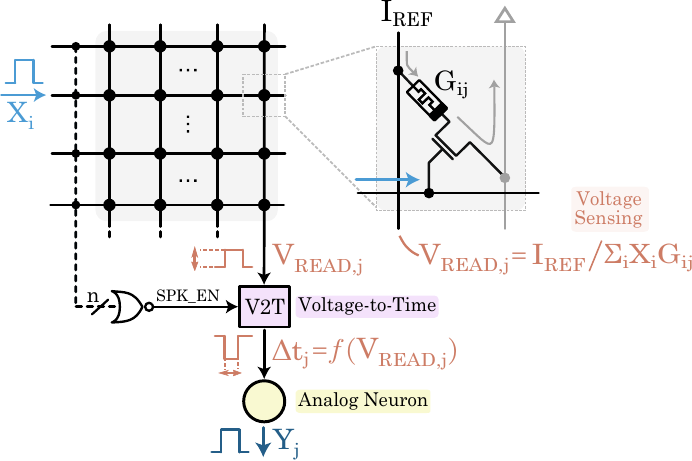}
\captionsetup{belowskip=-11pt}
\caption{Analog SNN with 1T1R memristive synapses employing voltage-mode crossbar sensing. The VMM output is represented by a column voltage $\subm{V}{READ,j}$ and converted into a pulse duration $\Delta\subm{t}{j}$ controlling the neuron current injection.} 
\label{fig:vmm-voltage}
\end{figure}

Fig.~\ref{fig:vmm-voltage} shows the proposed fully analog memristive SNN architecture. Spiking inputs and output neurons are connected through 1T1R synapses, with weights encoded in the conductance $\subm{G}{ij}$. \textcolor{black}{ The VMM output is the column conductance $\sum_{\text{i}} \subm{X}{i}\subm{G}{ij}$ (with $\subm{X}{i} \in \{0,1\}$), sensed as a voltage $\subm{V}{READ,j}$}. 

A reference current $\subm{I}{REF}$ is injected into each column, selected to keep $\subm{V}{READ,j}$ below the SET voltage. Memristors share a common top electrode, so the current is distributed among active devices, yielding:

\begin{equation}
    \subm{V}{READ,j} = \frac{\subm{I}{REF}}{\sum_{\text{i}} \subm{X}{i}\subm{G}{ij}}.
    \label{eq:vread}
\end{equation}

\noindent where the ON-resistance of the access transistor is neglected.

\begin{figure*}[t]
    \centering
    \begin{subfigure}{0.62\textwidth}
        \centering
        \includegraphics[width=\linewidth]{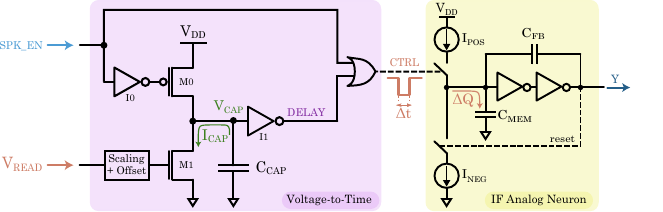}
        \caption{Conceptual schematic of the \myCircuit and the IF neuron. Capacitor discharge through an NMOS transistor implements the voltage-to-time conversion, while the OR gate produces the variable-width pulse that injects current into the neuron. For simplicity, the j subscript is omitted.}
        \label{fig:block-diagram}
    \end{subfigure}
    \hfill
    \begin{subfigure}{0.37\textwidth}
        \centering
        \includegraphics[width=\linewidth]{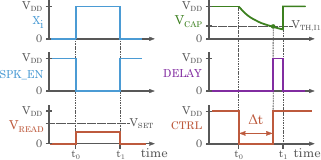}
        \caption{Representative waveforms showing the timing relationship between the input spike $\subm{X}{i}$, SPK\_EN, $\subm{V}{READ}$, $\subm{V}{CAP}$, DELAY, and CTRL for a fixed $\subm{V}{READ}$.  
        }
        \label{fig:waveforms}
    \end{subfigure}
    \caption{\myCircuitns: (a) high-level conceptual schematic including the IF neuron (b) illustrative waveforms.}
    \label{fig:delay-circuit}
\end{figure*}

The proposed voltage-to-time (V2T) readout circuit converts the column voltage into a variable-width pulse. The circuit receives $\subm{V}{READ,j}$ and a global enable SPK\_EN, which is low while any input spike is present. The output signal CTRL is an active-low pulse with duration $\Delta \subm{t}{j}$. Fig.~\ref{fig:delay-circuit} shows a high-level schematic and representative waveforms. 

The voltage-to-time conversion is implemented by modulating the discharge time of capacitor $\subm{C}{CAP}$. During SPK\_EN, capacitor $\subm{C}{CAP}$ discharges through NMOS M1. After SPK\_EN, PMOS M0 restores $\subm{V}{CAP}$. The discharge time required for $\subm{V}{CAP}$ to reach the inverter threshold $\subm{V}{TH,I1}$ is:

\begin{equation}
    \Delta \subm{t}{j} = \frac{\subm{C}{CAP} \cdot \Delta\subm{V}{CAP}}{\subm{I}{CAP,j}} 
\label{eq:discharge}
\end{equation}

\noindent with $\Delta\subm{V}{CAP} = \subm{V}{DD}-\subm{V}{TH,I1}$. Assuming transistor M1 operates in saturation, the capacitor discharge current $\subm{I}{CAP,j}$ follows the alpha-power-law. To ensure saturation, M1 is driven by a scaled and offset version of $\subm{V}{READ,j}$ so that $\subm{V}{GS,j} = \subm{k}{OF} \cdot \subm{V}{READ,j} + \subm{V}{OF}$. Therefore, $\subm{I}{CAP,j}$ can be expressed as:

\begin{equation}
    \subm{I}{CAP,j} \propto ((\subm{k}{OF} \cdot \subm{V}{READ,j} + \subm{V}{OF})-\subm{V}{T})^\alpha
\label{eq:ICAP}
\end{equation}

Combining \eqref{eq:vread}, \eqref{eq:discharge} and \eqref{eq:ICAP}, the relation between the pulse duration $\Delta \subm{t}{j}$ and the column conductance $\sum_{\text{i}} \subm{X}{i}\subm{G}{ij}$ is obtained:

\begin{equation}
    \Delta\subm{t}{j} \propto \frac{\subm{C}{CAP} \cdot \Delta\subm{V}{CAP}}{ \left((\subm{k}{OF} \cdot \cfrac{\subm{I}{REF}}{\sum_{\text{i}}\subm{X}{i}\subm{G}{ij}} + \subm{V}{OF})-\subm{V}{T} \right) ^\alpha} 
\label{eq:discharge-full}
\end{equation}

Equation~\eqref{eq:discharge-full} describes the resulting conductance-to-time conversion, which is monotonic but inherently non-linear, following a power-law dependence of $\Delta \subm{t}{j}$~on $\sum_{\text{i}} \subm{X}{i}\subm{G}{ij}$. Through circuit tuning, a locally approximately linear region can be achieved, as shown in the next section.

CTRL is generated by OR-ing SPK\_EN with the inverted capacitor voltage. As a result, the capacitor discharge time $\Delta \subm{t}{j}$ sets the pulse width (neglecting gate delays).

The IF neuron (Fig.~\ref{fig:block-diagram}) incorporates a switched current source $\subm{I}{POS}$ controlled by CTRL. The charge injected into the neuron is therefore proportional to the pulse duration $\Delta\subm{t}{j}$:

\begin{equation}
    \Delta\subm{Q}{j} = \subm{I}{POS}\cdot\Delta\subm{t}{j}
\label{eq:charge}
\end{equation}

Since the VMM output is encoded in \textit{time}, $\subm{I}{POS}$ can be set by neuron design constraints. This contrasts with current-sensing schemes, where the VMM result is encoded in the current \textit{amplitude} and scales with the column output.


\section{Results and Discussion}
\label{sec:results}
\begin{figure*}[!htbp]
\centering
\includegraphics[width=\textwidth]{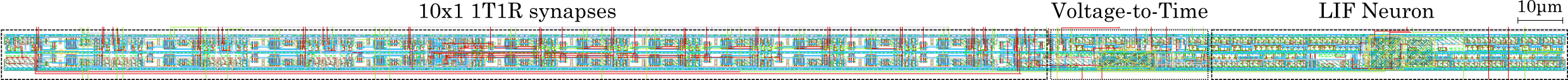}
\captionsetup{belowskip=-9pt}
\caption{Layout of the implemented $10 \PLH 1$ SNN prototype. Total area is \qty{2880.52}{\micro\metre\squared}.}
\label{fig:chip-layout}
\end{figure*}

A $10 \PLH 1$ SNN prototype based on the architecture in Fig.~\ref{fig:vmm-voltage} was implemented in \qty{130}{\nano\metre} CMOS and submitted for fabrication. The layout is shown in Fig.~\ref{fig:chip-layout} and design parameters are summarized in Table~\ref{tab:design-targets}. 

\begin{table}[!htbp]
\renewcommand{\arraystretch}{1.2}
\centering
\setlength{\tabcolsep}{8pt}
\caption{Design parameters for the $10 \PLH 1$ SNN prototype}
\label{tab:design-targets}
\begin{tabular}{lll}
\hline
\textbf{Parameter} & \textbf{Value }& \textbf{Description} \\
\hline
$\subm{G}{ij}$   & \qtyrange[range-phrase=--]{0.1}{1}{\milli\siemens}   & Memristor conductance \\
$W/L$            & \qty{3.9}{\micro\meter}/\qty{0.13}{\micro\meter}     & 1T1R access transistor \\
$\subm{V}{READ}$ & \qtyrange[range-phrase=--]{2.5}{250}{\milli\volt}    & Column voltage \\
$\subm{I}{REF}$  & \qty{25}{\micro\ampere}                              & Column current reference \\
$\subm{C}{CAP}$  & \qty{60}{\femto\farad}                               & V2T capacitor \\
$\subm{C}{MEM}$  & \qty{66.5}{\femto\farad}                             & Neuron membrane capacitor \\
$\subm{C}{FB}$   & \qty{47.5}{\femto\farad}                             & Neuron feedback capacitor \\
$\subm{I}{POS}$  & \qty{25}{\nano\ampere}                               & Neuron charging current \\
$\subm{I}{NEG}$  & \qty{40}{\nano\ampere}                               & Neuron discharging current \\
\hline
\end{tabular}
\end{table}

The \myCircuit is designed for input spikes with \qty{2.5}{\micro\second} pulse width at rates up to \qty{200}{\kilo\hertz}. The scaling and offset network is implemented using diode-connected MOS transistors, sized to draw negligible current compared to active memristors so that the read operation is not disturbed. The IF neuron operates on the input spikes timescale. All current sources are implemented using conventional CMOS current source structures. \textcolor{black}{Robustness was verified through PVT corners and Monte Carlo simulations, with $\sigma/\mu=$\qty{13.7}{\percent} timing spread at the maximum-conductance operating point under combined process and mismatch variations.}

\subsection{Post-layout simulation results}
\label{sec:post-layout-results}
This subsection presents post-layout simulation results of the $10 \PLH 1$ SNN prototype. The memristive devices are simulated using the Verilog-A model available in the PDK \cite{ihp}.

\begin{figure}[!htbp]
\centering
\includegraphics[width=0.9\columnwidth]{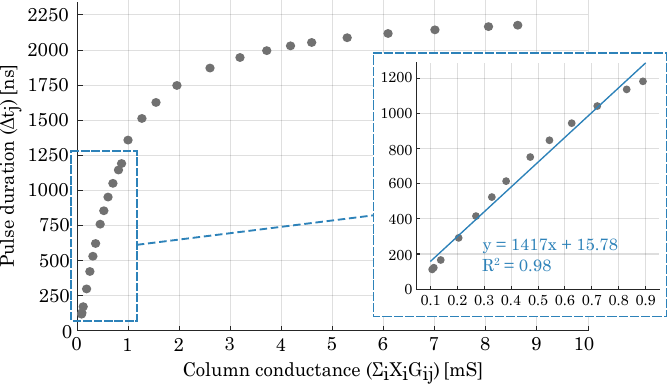}
\captionsetup{belowskip=-12pt}
\caption{Post-layout simulation results of the conductance-to-time conversion characteristic of the $10 \PLH 1$ SNN prototype.} 
\label{fig:delay-vs-G}
\end{figure}

Fig.~\ref{fig:delay-vs-G} shows the pulse duration $\Delta \subm{t}{j}$ as a function of the column conductance $\sum_{\text{i}} \subm{X}{i}\subm{G}{ij}$, obtained from a representative subset of active inputs and conductance combinations. From Fig.~\ref{fig:delay-vs-G}, it is confirmed that the relation in \eqref{eq:discharge-full} is monotonic, enabling time-based representation of the VMM output over the operating range. \textcolor{black}{The inset shows the approximately linear region, while saturation at higher conductance reflects the underlying power-law dependence. At high column conductance, additional simultaneously active inputs reduce the sensitivity of the conversion rather than preventing circuit operation.}

\begin{figure*}[t]
\centering
\includegraphics[width=0.85\textwidth]{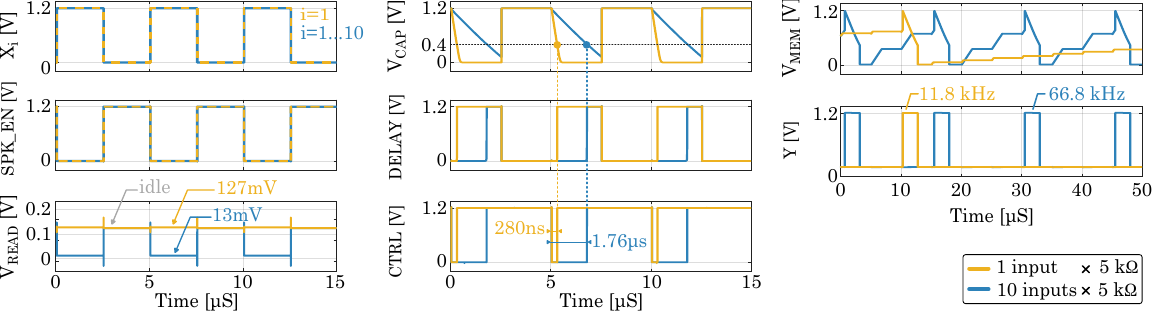}
\captionsetup{belowskip=-11pt}
\caption{Post-layout simulation waveforms of the $10 \PLH 1$ SNN prototype, illustrating two scenarios: one (orange) and ten (blue) simultaneously active inputs at \qty{200}{\kilo\hertz}, corresponding to column conductances of \qty{0.2}{\milli\siemens} and \qty{2}{\milli\siemens}, respectively.} 
\label{fig:results}
\end{figure*}

Fig.~\ref{fig:results} shows representative circuit waveforms for one and ten simultaneously active inputs at \qty{200}{\kilo\hertz} and \qty{0.2}{\milli\siemens} synaptic weights. The resulting column voltage modulation produces distinct CTRL pulse durations, leading to neuron firing rates of \qty{11.8}{\kilo\hertz} and \qty{66.8}{\kilo\hertz}, respectively.

\subsection{Area and Energy}
Table~\ref{tab:area} reports the layout area of the implemented prototype (excluding access circuitry), also expressed in equivalent minimum-size inverters (\qty{7.56}{\micro\metre\squared}).

\begin{table}[!htbp]
\renewcommand{\arraystretch}{1.2}
\centering
\setlength{\tabcolsep}{10pt}
\caption{Area of the Implemented Prototype}
\label{tab:area}
\begin{tabular}{lll}
\hline
\textbf{Sub-circuit} & \textbf{Area [\unit{\micro\metre\squared}]} & \textbf{Area [INV]} \\
\hline
$10 \PLH 1$ 1T1R synapses  & \num{393}     & \num{52}\\
\myCircuit                   & \num{480}     & \num{63.5} \\
IF Neuron                   & \num{1160}    & \num{153.5} \\
\hline
\textbf{Total }              & \num{2033}    & \num{269} \\
\hline
\end{tabular}
\end{table}

Energy is evaluated at the level of the synaptic circuit \textcolor{black}{(including the reference current generator)} and the \myCircuitns. 
The $10 \PLH 1$ synaptic circuit draws \qty{42.8}{\micro\ampere} independently of input activity, resulting in \qty{128.5}{\pico\joule} per read operation over \qty{2.5}{\micro\second}. The \myCircuit draws \qty{10.3}{\micro\ampere} in idle and \qty{0.19}{\micro\ampere} during conversion, yielding \qty{0.58}{\pico\joule} per operation. The total energy per synaptic operation (synaptic readout plus voltage-to-time conversion) is \qty{129}{\pico\joule}.

Area and \textcolor{black}{idle} power scale \textcolor{black}{approximately} linearly with the number of output neurons, \textcolor{black}
{as each requires one $\subm{I}{REF}$ current generator and one V2T circuit.} However, $\sim$70\% of the \myCircuit area is occupied by decoupling capacitors, which can be shared in larger systems. Idle power can be reduced by disabling the $\subm{I}{REF}$ current generator when no input is active. 

\vspace{-0.25em}

\subsection{Comparison with a current-sensing scheme}

As a representative example of a current-sensing readout, \cite{Garg2024} employs an opamp-based input stage and a current-scaling circuit providing $500\times$ attenuation. The per-column readout area in \qty{130}{\nano\metre} CMOS, estimated from an optical micrograph, occupies \qty{11000}{\micro\metre\squared} ($\approx 1455$ INV), over $20\times$ larger than the proposed \myCircuitns.

The power per read operation reported in \cite{Garg2024} is \qty{48}{\micro\watt}, corresponding to \qty{120}{\pico\joule} over our \qty{2.5}{\micro\second} time window. While this energy is of the same order of magnitude, \cite{Garg2024} considers a current range of \qtyrange[range-phrase=--]{10}{200}{\micro\ampere}. These currents correspond to a column conductance of \qtyrange[range-phrase=--]{0.04}{0.8}{\milli\siemens} (at \qty{250}{\milli\volt}), which is $<10\%$ of the \qtyrange[range-phrase=--]{0.1}{10}{\milli\siemens} range targeted here.

Using a current-sensing scheme, our $10 \PLH 1$ SNN would produce column currents from \qty{25}{\micro\ampere} to \qty{2.5}{\milli\ampere}. Scaling these to the \unit{\nano\ampere}-range would require attenuation factors of \num{e3}--\num{e5} and additional circuitry to enforce a fixed read voltage, which would increase the energy of \qty{120}{\pico\joule} by several orders.


\vspace{-0.25em}

\subsection{Application to Digit Recognition}

To further evaluate the proposed approach, a $64\PLH10$ SNN was trained in software on the Digit dataset \cite{scikit-digit} using back-propagation, with synaptic weights constrained to be non-negative. The trained weights were quantized to ten conductance levels spanning the memristor range (\qtyrange{1}{10}{\kilo\ohm}) and mapped to a schematic-level implementation, where the IF analog neurons were extended with leakage (implemented as a constant current source) and lateral inhibition.

\begin{figure}[!htbp]
\centering
\includegraphics[width=\columnwidth]{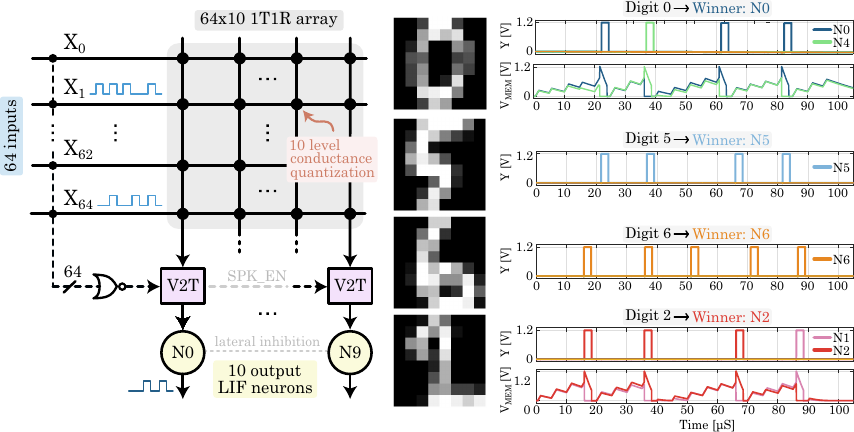}
\captionsetup{belowskip=-9pt}
\caption{Application of the proposed architecture to a trained $64\PLH10$ SNN with quantized synaptic weights. Membrane traces are included only when multiple neurons are active.} 
\label{fig:64x10}
\end{figure}

Representative $8\PLH8$ digit samples, rate-encoded over \qty{100}{\micro\second}, produced the same classification outcome in both the software-trained network and the circuit implementation (Fig.~\ref{fig:64x10}), providing preliminary validation of the proposed architecture in a trained SNN inference task.

\vspace{-0.7em}

\section{Conclusion}
This work proposes a time-based readout circuit for fully analog SNNs with memristive synapses. The proposed approach converts the VMM output, sensed as a column voltage, into a time-domain signal that controls neuron current injection. By avoiding current-mode readout, the approach eliminates complex, area-intensive circuitry that becomes challenging at high column currents.

A $10\PLH1$ SNN prototype with the proposed architecture was implemented in \qty{130}{\nano\meter} CMOS. Post-layout results confirm correct conductance-to-time conversion, achieving $\sim$20$\PLH$ area reduction over a representative current-sensing implementation with comparable energy efficiency. Application to a trained $64\PLH10$ SNN further demonstrates the feasibility of the architecture for inference with quantized memristive weights. 

\vspace{-0.4em}

\section*{Acknowledgment}
This work has been supported by PID2022-141391OB-C22 funded by MCIN/AEI/10.13039/501100011033/FEDER, UE. The corresponding author gratefully acknowledges the Universitat Politècnica de Catalunya and Banco Santander for the financial support of her predoctoral FPI-UPC grant.

\vspace{-0.4em}



\bibliographystyle{IEEEtran}
\bibliography{papers_overleaf, additional_bib}

\end{document}